\documentclass[aps,showpacs,prl,twocolumn]{revtex4}
\usepackage{graphicx}

\begin{document}

\title{Calorimetric Evidence for Nodes in the Overdoped Ba(Fe$_{0.9}$Co$_{0.1}$)$_{2}$As$_{2}$}

\author{Dong-Jin Jang$^1$, A. B. Vorontsov$^2$, I. Vekhter$^3$, K. Gofryk$^4$, Z. Yang$^1$, S. Ju$^1$, J. B. Hong$^1$, J. H. Han$^1$, Y. S. Kwon$^1$, F. Ronning$^4$, J. D. Thompson$^4$, and T. Park$^{1,4}$}

\address{$^1$Department of Physics, Sungkyunkwan University, Suwon 440-746, Republic of Korea \\ $^2$ Department of Physics, Montana State University, Bozeman, Montan 59717, USA \\$^3$ Department of Physics and Astronomy, Louisiana State University, Baton Rouge, LA 70803-4001, USA \\ $^4$Los Alamos National Laboratory, Los Alamos, NM 87545, USA }

\begin{abstract}
We present low-temperature specific heat of the electron-doped Ba(Fe$_{0.9}$Co$_{0.1}$)$_{2}$As$_{2}$, which does not show any indication of an upturn down to 400~mK, the lowest measuring temperature. The lack of a Schottky-like feature at low temperatures or in magnetic fields up to 9~Tesla enables us to identify enhanced low-temperature quasiparticle excitations and to study anisotropy in the linear term of the specific heat. Our results can not be explained by a single or multiple isotropic superconducting gap, but are consistent with multi-gap superconductivity with nodes on at least one Fermi surface sheet. 
\end{abstract}

\maketitle

Since the discovery of the iron-based pnictide superconductors,
substantial experimental and theoretical work has been performed to
understand the superconducting (SC) mechanism and its manifestation via the gap structure.~\cite{Ishida} Despite these efforts, however, the gap symmetry is still unclear. In RFeAsO (R: rare earth), the '1111' phase, a full gap is expected from Andreev spectroscopy,~\cite{Chen} but a nodal gap is reported from optical measurements~\cite{Dubroka} for a compound with similar stoichiometry. Conflicting interpretations of the SC order parameter also have been reported in Co-doped BaFe$_2$As$_2$ (Ba122 phase). Angle-resolved photoemission spectroscopy
(ARPES)~\cite{Terashima} and scanning tunneling microscopy (STM)~\cite{Yin} suggest an isotropic gap; whereas, Raman scattering,~\cite{Mushuler} nuclear magnetic resonance (NMR)~\cite{Ning}, muon spin rotation ($\mu$SR)~\cite{Williams}, thermal conductivity~\cite{Tanatar,Dong}, specific heat,~\cite{Mu0} and penetration depth~\cite{Gordon,hashimoto} measurements all indicate an anisotropic gap on one or more Fermi surface sheets. Strong doping dependence of superconducting properties and/or specific features of the measurement techniques which probe preferentially a fraction of the total Fermi surface could be behind these seemingly contradicting results.~\cite{McQueen, Kim0}

Low-temperature specific heat measurements, which probe the electronic density of states, have been instrumental in determining the SC order parameter of various classes of superconductors, including high-$T_c$ cuprates.~\cite{moler94, park03, sakakibara07} In the newly discovered Fe-based pnictides, the interpretation of specific heat measurements at $T \ll T_c$ has been controversial because of a Schottky-like upturn at the lowest temperatures, which has been attributed to some kind of impurities.~\cite{McQueen, Gofryk, Mu, Kim} Here we report low-temperature specific heat measurements of Ba(Fe$_{0.9}$Co$_{0.1}$)$_{2}$As$_{2}$, in which there is no Schottky-like feature down to 400~mK and up to 9~T, the lowest measured temperature and the highest applied magnetic field, respectively. The lack of Schottky contribution allows us to identify the low-$T$ electronic excitations and to study the field dependence of the linear specific heat term. Our results are not compatible with an isotropic s-wave gap, but are consistent with gap zeroes on parts of a Fermi surface in a multiband model. 

Single crystalline Ba(Fe$_{0.9}$Co$_{0.1}$)$_{2}$As$_{2}$ was grown
using a closed Bridgman method with a tungsten crucible at 1550~$^0$C to contain volatile arsenic (As)~\cite{song10}. Powder x-ray resonance diffraction shows that the lattice parameters are a=3.9458~$\AA$ and c=12.9476~$\AA$. These crystals have a plate-like tendency with the [001]-direction perpendicular to the crystal plane. Specific heat was measured down to 400~mK and up to 9~T in a Quantum Design PPMS (Physical Properties Measurement System) with a $^3$He option.  The superconducting transition temperature, defined from the midpoint of the specific heat discontinuity (see Fig.~2a), is 19.0~K, which is lower than that of optimally doped Ba(Fe$_{0.92}$Co$_{0.08}$)$_{2}$As$_{2}$ ($T_c=25$~K) and indicates that the present compound is slightly overdoped. 

Figure~1 is a plot of the temperature-dependent specific heat of
Ba(Fe$_{0.9}$Co$_{0.1}$)$_{2}$As$_{2}$ in zero field. A specific heat jump at 19.0~K signals the superconducting phase transition, confirmed by low-field magnetic susceptibility measurements. Unlike magnetic and structural transitions in Ba(Fe$_{1-x}$Co$_{x}$)$_{2}$As$_{2}$, which are strongly affected by Co-doping, the phononic contribution to the specific heat is almost independent of doping.~\cite{Ni, Chu} In the ensuing analysis, therefore, we use the phonon specific heat measured from undoped BaFe$_{2}$As$_{2}$ as the phonon contribution in the present material. Because there is no Schottky-like feature at the lowest temperatures of these measurements, the normal-state specific heat ($C_n$) can be written as $C_n/T=\gamma_n + C_{ph}/T$, where $\gamma_n$ is the normal-state Sommerfeld coefficient and $C_{ph}$ is the phonon contribution. The solid curve in Fig.~1 is the normal state $C_n/T$ with $\gamma_n$=23.5 mJ/mol$\cdot$K$^{2}$, which satisfies entropy conservation at $T_c$ (see the inset of Fig.~1 and discussion below). In the superconducting state, there is a noticeable $T-$linear specific heat contribution $\gamma_0$ of 3.4~mJ/mol$\cdot$K$^{2}$, about 14~$\%$ of the normal-state $\gamma_n$. A sizable $\gamma_0$ is often reported in superconductors with gap zeroes on parts of the Fermi surface because impurities easily produce a finite density of states even at zero energy. This finite value of $\gamma_0$ suggests the possibility of nodal superconductivity in Ba(Fe$_{0.9}$Co$_{0.1}$)$_{2}$As$_{2}$. Even though a finite $\gamma_0$ could arise from non-superconducting parts of the crystal, the lack of a Schottky-like feature in $C/T$ and lack of a second impurity phase in x-ray diffraction suggest that any contribution from impurities is negligibile in the ensuing analysis. Supporting this conclusion, the value of $\gamma_0$ is smaller than that reported earlier on a compound with similar Co concentration and an upturn in the low$-T$ specific heat~\cite{gofryk10}. We note that in multiband systems, a zero-energy impurity band can be effectively formed without nodes on any Fermi surface sheet, resulting in large value of $\gamma_0$~\cite{bang10}.
\begin{figure}
\centering  \includegraphics[width=7cm,clip]{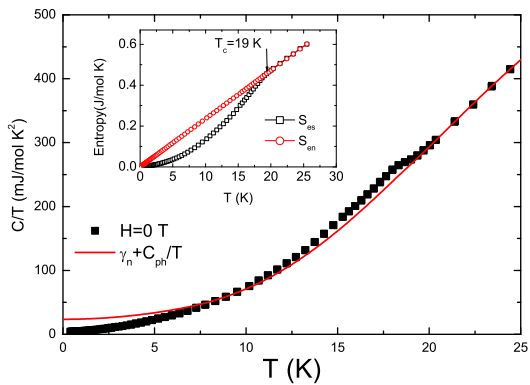}
\caption{(color online) Specific heat divided by temperature of
Ba(Fe$_{0.9}$Co$_{0.1}$)$_{2}$As$_{2}$ in zero applied field (filled squares). The red solid line is the phonon contribution $C_{ph}/T$ determined from the parent compound BaFe$_2$As$_2$, plus a constant $\gamma_n$=23.5 mJ/mol K$^2$ that is required to balance entropy at $T_c$. Inset: Electronic entropy of normal (red circles) and  superconducting (black squares) states. A deviation of $S_{es}$ from $S_{en}$ appears near 19~K.}
\label{figure1}
\end{figure}

Electronic entropies of the normal ($S_{en}$) and superconducting ($S_{es}$) phases of Ba(Fe$_{0.9}$Co$_{0.1}$)$_2$As$_2$ are plotted in the inset of Fig.~1 and are used to estimate various SC properties. The superconducting condensation energy, $U=\int_{0}^{T_c}(S_{en}-S_{es})dT$, is estimated to be 1.23~J/mol. The zero-temperature thermodynamic critical field $B_c(0)$ is 0.22~T, from the relation $U=B_c^2(0)/2\mu_0$. When combined with previous measurements of the penetration depth ($\lambda=325$~nm) and coherence length ($\xi=27.6$~nm),~\cite{Yin,Luan,Bernhard} the upper critical field
is estimated to be 37~T from the relation $\mu_0H_{c2}=\sqrt{2}\kappa B_c(0)$, $\kappa=\lambda/\xi=11.8$. These superconducting properties of the present sample with $x=0.10$ are quite comparable to those reported previously on compounds with similar stoichiometry.~\cite{Gofryk}  

Figure~2 displays the low-temperature electronic specific heat of Ba(Fe$_{0.9}$Co$_{0.1}$)$_2$As$_2$ after subtracting the phononic contribution. As seen in Fig.~2b, there is no hint of an upturn in $C_{el}/T$ down to 400~mK and up to 9~T, irrespective of the direction of the applied field (see Fig.~3). Thus the closed Bridgman method used for preparing this crystal seems to avoid incorporation of impurity phases that are generated from a flux-growth method, enabling us to study the low-temperature specific heat without the need to subtract poorly understood values to account for the field-dependent upturn in previously studied samples. 
\begin{figure}
\centering  \includegraphics[width=7cm,clip]{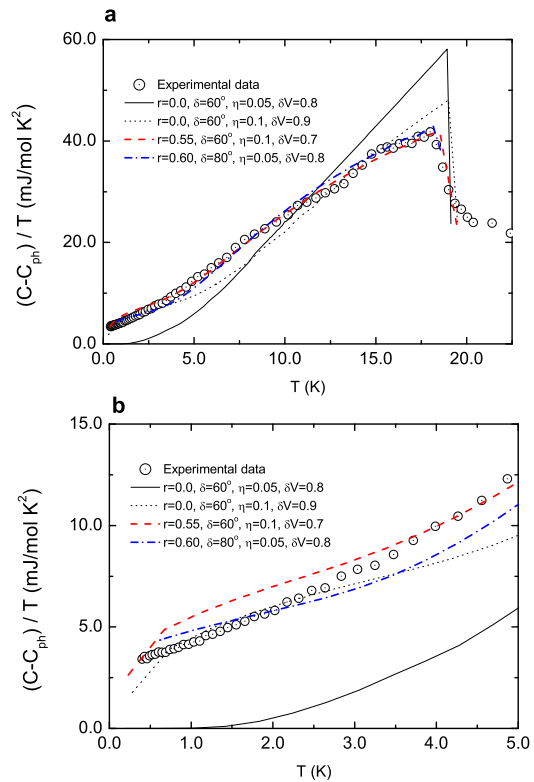}
\caption{(color online) \textbf{a} Temperature dependence of the electronic specific heat. The experimental data are compared with the two-band model described in the text, with an isotropic gap on the hole sheet, and a nodal gap on the electron sheet. Note that, when the nodal behavior is weak ($r=0.55$, nodes close to each other), the numerical result using the scattering rates required to explain the residual linear term in the superconducting state exhibits a downturn at low temperatures, not observed experimentally. Therefore the low-temperature data are most consistent with a relatively well-pronounced nodal behavior, $r\geq 0.6$. \textbf{b} Magnification of the electronic specific heat at low temperatures.}
\label{figure2}
\end{figure}

To investigate the gap symmetry, we begin by analyzing the temperature dependence of $C_{el}/T$ in zero field, where $C_{el}=C-C_{ph}$. We analyze the data using an effective two band model with an isotropic gap on the hole Fermi surface and a gap on the electron Fermi surface of the form $\Delta_e( \textbf{k})=\Delta_e (1-r+r\cos 2\phi)$, where $\phi$ is the angle measured from the [100] direction. This form is suggested by several theories~\cite{chubukov10, maier09}, and is consistent with the general $A_{1g}$ symmetry of the order parameter in the Brillouin Zone. In this expression for the gap function, $r>0.5$ ($r<0.5$) corresponds to the nodal (fully gapped) order parameter on the electron Fermi surface sheet. The gap magnitudes on the two Fermi surfaces were computed self-consistently assuming a purely interband pairing interaction requiring a sign change of the order parameter between the electron and the hole sheets~\cite{mazin08, chubukov08}. The residual linear, in $T$, term and the temperature variation of the electronic specific heat are controlled by the value of $r$ and by disorder. In our model the impurities are characterized by their concentration, $n_{imp}$, and the strength of the inter- and intra-band scattering, $U_1$ and $U_0$ respectively. The quasiparticle scattering rate $\Gamma$ depends also on the density of states at the Fermi surface, $N_f$, which , for simplicity, we take to be the same on the two sheets. We use the transition temperature of the pure (impurity-free) sample, $T_{c0}$, as a unit of energy, and hence introduce a dimensionless scattering rate $\eta=\Gamma_n / 2\pi T_{c0}$, where $\Gamma_n = n_{imp}/\pi N_f$. Together with the phase shift of the intraband scattering, $\delta=arctan(\pi N_f U_0)$, and the ratio of the inter- to intra-band potential $\delta V = U_1 / U_0$ it completely determines the properties of the impurity ensemble. Since the inter- (intra-) band potential describes scattering with a large (small) momentum transfer, for usual charged impurities we expect $\delta V \leq1$, although $\delta V \approx 1$ is plausible for the direct substitution to the Fe site that strongly affects the $d$ orbitals. Indeed, our best fits give a moderate ratio $\delta V$. We determine the quasiparticle self-energies using the self-consistent $T-$matrix approximation~\cite{hirschfeld1988,balatsky2006,mishra09}, and compute the density of states, the entropy, and the specific heat for comparison with the experimental results.

\begin{figure}
\centering  \includegraphics[width=7cm,clip] {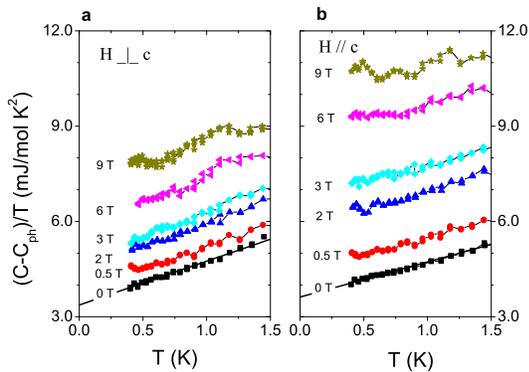}
\caption{(color online) Low-temperature specific heat plotted for various magnetic fields applied parallel and perpendicular to the crystalline $c$-axis in panel \textbf{a} and \textbf{b}, respectively. Solid lines at zero field are guides to eyes that show a $C\propto T^2$ dependence.}
\label{figure3}
\end{figure} 
We found that, irrespective of the strength of the scattering and the impurity concentration, the low-temperature data cannot be reliably fit using a model with a fully gapped electron Fermi surface, $r<0.5$. Even though the density of states at the Fermi surface, and therefore the Sommerfeld coefficient $\lim_{T\rightarrow 0} C_{el}(T)/T$, may be finite in this case due to formation of an impurity band in the presence of strong interband scattering~\cite{mishra09}, more curvature is expected for the temperature dependence of $C_{el}$ in that case than is observed experimentally. The nearly linear dependence in Fig.~2 can only be satisfactorily explained assuming a nodal gap structure, i.e., $r>0.5$. Even in that case, the intraband scattering tends to lift the nodes~\cite{mishra09}, resulting in a pronounced downturn in $C_{el}/T$ for the values of $r$ close to the ``critical'' value that separates the gapless regime from that where the nodes are lifted by disorder and a full gap opens (see Fig.~2). As shown in Fig.~2, the best fits are for the well-defined nodes, $r=0.6$.
\begin{figure}
\centering  \includegraphics[width=7cm,clip]{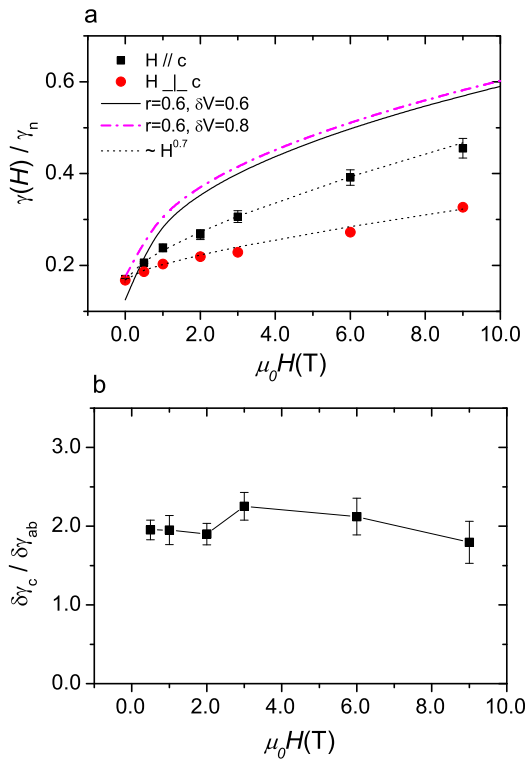}
\caption{(color online) \textbf{a} Field dependence of the linear term in the heat capacity at 0.4~K ($T/T_c=0.021$) divided by the normal state Sommerfeld coefficient $\gamma_n$. Theoretical calculations using the same parameters that give the best fit to the temperature dependence of $C_{el}/T$ in zero field reproduce the overall field dependence but yield a higher magnitude of the effect: the dashed line is calculated from the parameters ($r=0.6, \delta=80^o, \eta=0.1, \delta V=0.6$), while the dash-dotted line is from ($r=0.6, \delta=80^o, \eta=0.05, \delta V=0.8$). As explained in text, the larger absolute value of the calculated $\gamma (H)$ is due to an overestimate of the contribution of the hole band within the modified Brandt-Pesch-Tewordt approximation. For reference, we show least-squares fits of the data for $H \parallel c$ and $H\perp c$ to a power law form $a+bH^{\epsilon}$ (dotted lines), where $n=0.7$ reasonably explains the data. \textbf{b} Anisotropic field dependence of the linear term in the heat capacity at 0.4~K. The ratio of the Sommerfeld coefficient for field parallel and perpendicular to the crystalline c-axis, $\delta \gamma_c / \delta \gamma_{ab}$, is plotted against magnetic field, where $\delta \gamma = \gamma(H) - \gamma(0 T)$}
\label{figure4}
\end{figure}

Measurements under applied magnetic field provide additional information on the gap structure. The electronic specific heat divided by temperature $C_{el}/T$ is displayed against $T$ for magnetic field perpendicular and parallel to the crystalline c-axis in Fig.~3a and 3b, respectively. With increasing field, the temperature variation of $C_{el}/T$ at low $T$ becomes weaker and almost disappears at 9~T for $H//c$. Since the extrapolated $T=0$ intercept of $C_{el}/T$, $\gamma_0$, depends on the power law that we use for the extrapolation, in Fig.~4a we plot the normalized values of the $C/T$ at 0.4~K ($T/T_c = 0.021$) to study the magnetic field dependence of the density of states. For an s-wave SC gap, the electronic density of states is proportional to the density of vortices in the mixed state, leading to a linear field dependence of $\gamma_0$. In contrast, the presence of nodes in the SC gap changes the field-dependence to a sub-linear behavior due to a Doppler shift of the nodal quasiparticles. Supporting the conclusion of nodal superconductivity from the low-$T$ specific heat analysis, $C_{el}/T$ which is proportional to the electronic density of states deviates from a linear-dependence on the field, but shows a sub-linear behavior. Comparison between the experimental data for field along the c-axis and the theory with SC gap nodes is shown in Fig.~4a, where we use the values of $r$ along with the impurity parameters from Fig.~2 to compute the field dependence of the specific heat under the applied field using the modified Brandt-Pesch-Tewordt (BPT) approximation~\cite{anton07}. This approximation is designed for intermediate and high fields and is known to give only a qualitatively correct picture for nodal gaps at low $H$. As seen in Fig.~4a, the calculations reproduce the sublinear field dependence, consistent with the existence of nodes. At the same time the magnitude of the field enhancement is much greater in theory than in experiment. This is due to a significant overestimate of the contribution of the fully gapped hole Fermi surface to the density of states within the BPT approximation at low fields~\cite{anton07, pesch75}.

The field dependence of the specific heat coefficient ($\gamma (H)$) is fit by a power-law form $a+bH^{0.7}$ for both field directions with $b=0.64$ and 0.33 for $H\parallel c$ and $H\perp c$, respectively (dotted lines in Fig. 4a). Anisotropy of the field-induced quasiparticle density of states, $\delta \gamma_c/\delta \gamma_{ab}$, is plotted in Fig.~4b, where $\delta \gamma = \gamma (H) - \gamma(0 T)$. This anisotropy ratio is approximately two over the measured field range and reflects the superconducting anisotropy through $\delta \gamma_c/\delta \gamma_{ab} = (H_{c2}^{ab}/H_{c2}^{c})^\epsilon$~\cite{ichioka99}. When $\epsilon=0.7$ is used, the uppercritical field anisotropy ratio is approximately 2.7. This value is compatible with that estimated from the initial slope $(dH_{c2}/dT)$ at $T_c$ by the Werthamer-Helfand-Hohenburg model~\cite{whh66, yamamoto09}. Kano et al., however, reported that the $H_{c2}$ anisotropy ratio at $T=0$~K in a compound with similar stoichiometry is close to 1, where the difference has been interpreted as due to multi-band effects in the pnictides~\cite{kano09}. Indeed, there is substantial evidence for multiple electronic bands in the pnictides and their consequences for superconductivity. Studies of the Fermi surface find a distinct asymmetry between hole and electron bands: a circular hole Fermi surface is observed around the $\Gamma$ point, but an anisotropic electron Fermi surface exists around the M point~\cite{mazin08,Seo,Yi,Liu}. Recently, penetration-depth and thermal conductivity measurements have suggested that line nodes may exist on an electron sheet near the Brillouin zone center and that hole bands centered around the $\Gamma$ point are fully gapped~\cite{fletcher09,hicks09,yamashita09,hashimoto09,reid10}.

In conclusion, the specific heat of Ba(Fe$_{0.9}$Co$_{0.1}$)$_{2}$As$_{2}$ prepared by the closed Bridgman method does not show any hint of an upturn at low temperatures, a materials problem that has plagued crystals prepared by other methods. The lack of a Schottky-like feature enables us to obtain the intrinsic low-temperature specific heat whose temperature dependence is consistent with an order parameter that produces an isotropic SC gap on one of the Fermi surface sheets and a gap with line nodes on the second, presumably electron, sheet. Evidence for nodal superconductivity is supported by the magnetic field dependence of the Sommerfeld coefficient $\gamma$ and a finite residual density of states in the superconducting state. Additionally, these experiments suggest that a complete description of superconductivity requires the involvement of multiple bands.  

This work was supported by the National Research Foundation (NRF) grant (2010-0016560) funded by Korea government (MEST). Work at Los Alamos was performed under the auspices of the U. S. Department of Energy/Office of Science and supported in part by the Los Alamos LDRD program. ZY and TP acknowledges support by the promotion program for new faculty, Sungkyunkwan University (2009). I.V. is supported in part by US DOE via Grant DE-FG02-08ER46492. ABV's support in part comes from US NSF grant DMR-0954342. Y.S.K. is supported by the Basic Science Research Program (2010-0007487) and Nuclear R\&D Programs (2006-2002165 and 2009-0078025).


\begin{thebibliography}{10}
\bibitem{Ishida} K. Ishida, Y. Nakai, and H. Hosono, J. Phys. Soc. Jpn. {\bf 78}, 062001 (2009).
\bibitem{Chen} T. Y. Chen, Z. Tesanovic, T. H. Liu, X. H. Chen, and C. L. Chien, Nature {\bf 453}, 1224 (2008).
\bibitem{Dubroka} A. Dubroka et al., Phys. Rev. Lett. {\bf 101}, 097011 (2008).
\bibitem{Terashima} K. Terashima et al., Proc. Natl. Acad. Sci. USA {\bf 106} 7330 (2009).
\bibitem{Yin} Y. Yin et al., Phy. Rev. Lett. {\bf 101}, 097002 (2009).
\bibitem{Mushuler} B. Mushuler et al., Phys. Rev. {\bf B 80}, 180510(R) (2009).
\bibitem{Ning} F. Ning et al., J. Phys. Soc. Jpn. {\bf 77}, 103705
(2008).
\bibitem{Williams} T. J. Williams et al., Phys. Rev. {\bf B 80}, 094501 (2009).
\bibitem{Tanatar} M. A. Tanatar et al., Phys. Rev. Lett. {\bf 104}, 067002 (2010).
\bibitem{Dong} J. K. Dong et al., Phys. Rev. {\bf B 81}, 094520 (2010).
\bibitem{Mu0} Gang Mu et al., Chin. Phys. Lett. {\bf 27}, 037402
(2010).
\bibitem{Gordon} R. T. Gordon et al., Phys. Rev. Lett. {\bf 102}, 127004 (2009).
\bibitem{hashimoto} K. Hashimoto et al., Phys. Rev. B {\bf 81}, 220501(R) (2010).
\bibitem{McQueen} T. M. McQueen et al., Phys. Rev. {\bf B 79}, 014522(2009).
\bibitem{Kim0} J. S. Kim et al., unpublished (arXiv:1002.3355v1).
\bibitem{moler94} K. A. Moler, D. J. Baar, J. S. Urbach, R. Liang, W. N. Hardy, and A. Kapitulnik, Phys. Rev. Lett. {\bf73}, 2744 (1994).
\bibitem{park03} T. Park et al., Phys. Rev. Lett. {\bf 90}, 177001 (2003). 
\bibitem{sakakibara07} T. Sakakibara et al., J. Phys. Soc. Jpn. {\bf 76}, 051004 (2007).
\bibitem{Gofryk} K. Gofryk et al., New. J. Phys. {\bf 12}, 023006 (2010).
\bibitem{Mu} Gang Mu et al., Phys. Rev. {\bf B 79}, 174501 (2009).
\bibitem{Kim} J. S. Kim, E. G. Kim and G. R. Stewart, J. Phys.: Condens. Matter {\bf 21}, 252201 (2009).
\bibitem{song10} Y. J. Song et al., Appl. Phys. Lett. {\bf 96}, 212508 (2010).
\bibitem{Ni} N. Ni et al., Phys. Rev. {\bf B 78}, 214515(R) (2008).
\bibitem{Chu} J.-H. Chu, J. Analytis, C. Kucharczyk, and I. R. Fisher, Phys. Rev. {\bf B 79}, 014506 (2009).
\bibitem{gofryk10} K. Gofryk et al., Phys. Rev. {\bf B 81}, 184518 (2010).
\bibitem{bang10} Y. Bang, Phys. Rev. Lett. {\bf 104}, 217001 (2010).
\bibitem{Luan} L. Luan et al., Phys. Rev. {\bf B 81}, 100501(R) (2010).
\bibitem{Bernhard} C. Bernhard et al., New J. Phys. {\bf 11}, 055050 (2009).
\bibitem{chubukov10} A. V. Chubukov and I. Eremin, Phys. Rev. B 82, 060504(R) (2010).
\bibitem{maier09} T. A. Maier, S. Graser, D. J. Scalapino, P. J. Hirschfeld, Phys. Rev. B 79, 224510 (2009).
\bibitem{mazin08} I. I. Mazin, D. J. Singh, M. D. Johannes, and M. H. Du, Phys. Rev. Lett. {\bf 101}, 057003 (2008).
\bibitem{chubukov08} A. V. Chubukov, D. V. Efremov, and I. Eremin, Phys. Rev. B {\bf 78}, 134512 (2008).
\bibitem{hirschfeld1988} P. J. Hirschfeld, P. W{\"o}lfle, and D. Einzel, Phys. Rev. B {\bf 37}, 83 (1988).
\bibitem{balatsky2006} A.~V.~Balatsky, I.~Vekhter, and J.-X.~Zhu, Rev. Mod. Phys. {\bf 78}, 373 (2006).
\bibitem{mishra09} V. Mishra, A. Vorontsov, P. J. Hirschfeld, and I. Vekhter, Phys. Rev. B 80, 224525 (2009).
\bibitem{anton07} A. B. Vorontsov and I. Vekhter, Phys. Rev. B 75, 224501 (2007).
\bibitem{pesch75} W. Pesch, Z. Phys. B 21, 263 (1975).
\bibitem{ichioka99} M. Ichioka, A. Hasegawa, and K. Machida, Phys. Rev. B {\bf 59}, 184 (1999).
\bibitem{whh66} N. R. Werthamer, E. Helfand, and P. C. Hohenberg, Phys. Rev. {\bf 147}, 295 (1966).
\bibitem{yamamoto09} Y. Mamoto et al., Appl. Phys. Lett. {\bf 94}, 062511 (2009).
\bibitem{kano09} M. Kano et al., J. Phys. Soc. Jpn. {\bf 78}, 084719 (2009).
\bibitem{Seo} K. Seo, B. A. Bernevig, and J. Hu, Phys. Rev. Lett.
{\bf 101}, 206404 (2008).
\bibitem{Yi} M. Yi et al., Phys. Rev. {\bf B 80}, 024515 (2009).
\bibitem{Liu} C. Liu et al. Phys. Rev. Lett. {\bf 101}, 177005 (2008).
\bibitem{fletcher09} J. D. Fletcher et al., Phys. Rev. Lett. {\bf102}, 147001 (2009).
\bibitem{hicks09} C. W. Hicks et al., Phys. Rev. Lett. {\bf103}, 127003 (2009).
\bibitem{yamashita09} M. Yamashita et al., Phys. Rev. B {\bf80}, 220509(R) (2009).
\bibitem{hashimoto09} K. Hashimoto et al., arXiv:0907.4399 (unpublished)
\bibitem{reid10} J.-Ph. Reid et al., Phys. Rev. B {\bf82}, 064501 (2010).
\end{thebibliography}
\end{document}